\documentclass[12pt, epsfig, twocolumn]{aastex6}
\usepackage{graphicx}
\usepackage{color}
\usepackage{epstopdf}

\shorttitle{Local Group distances and publication bias. V. Galactic
  rotation constants} 
\shortauthors{Richard de Grijs and Giuseppe Bono}

\begin{document}

\title{Clustering of Local Group distances: publication bias or
  correlated measurements? V. Galactic rotation constants}

\author{
Richard de Grijs\altaffilmark{1,2} and
Giuseppe Bono\altaffilmark{3,4}
}

\altaffiltext{1} {Kavli Institute for Astronomy \& Astrophysics and
  Department of Astronomy, Peking University, Yi He Yuan Lu 5, Hai
  Dian District, Beijing 100871, China}
\altaffiltext{2} {International Space Science Institute--Beijing, 1
  Nanertiao, Zhongguancun, Hai Dian District, Beijing 100190, China}
\altaffiltext{3} {Dipartimento di Fisica, Universit\`a di Roma Tor
  Vergata, via Della Ricerca Scientifica 1, 00133, Roma, Italy}
\altaffiltext{4} {INAF, Rome Astronomical Observatory, via Frascati
  33, 00040, Monte Porzio Catone, Italy}

\begin{abstract}
As part of on an extensive data mining effort, we have compiled a
database of 162 Galactic rotation speed measurements at $R_0$ (the
solar Galactocentric distance), $\Theta_0$. Published between 1927 and
2017 June, this represents the most comprehensive set of $\Theta_0$
values since the 1985 meta analysis that led to the last revision of
the International Astronomical Union's recommended Galactic rotation
constants. Although we do not find any compelling evidence of the
presence of `publication bias' in recent decades, we find clear
differences among the $\Theta_0$ values and the $\Theta_0/R_0$ ratios
resulting from the use of different tracer populations. Specifically,
young tracers (including OB and supergiant stars, masers, Cepheid
variables, H{\sc ii} regions, and young open clusters), as well as
kinematic measurements of Sgr A* near the Galactic Center, imply a
significantly larger Galactic rotation speed at the solar circle and a
higher $\Theta_0/R_0$ ratio (i.e., $\Theta_0 = 247 \pm 3$ km s$^{-1}$
and $\Theta_0/R_0 = 29.81 \pm 0.32$ km s$^{-1}$ kpc$^{-1}$;
statistical uncertainties only) than any of the tracers dominating the
Galaxy's mass budget (i.e., field stars and the H{\sc i}/CO
distributions). Using the latter as most representative of the bulk of
the Galaxy's matter distribution, we arrive at an updated set of
Galactic rotation constants,
\begin{eqnarray}
\Theta_0 &=& 225 \pm 3 \mbox{ (statistical)} \pm 10 \mbox{ (systematic)
  km s}^{-1},\nonumber \\
R_0 &=& 8.3 \pm 0.2 \mbox{ (statistical)} \pm 0.4 \mbox{ (systematic)
  kpc}, \mbox{ and}\nonumber \\
\Theta_0 / R_0 &=& 27.12 \pm 0.39 \mbox{ (statistical)} \pm 1.78
\mbox{ (systematic) km s}^{-1} \mbox{ kpc}^{-1}.\nonumber
\end{eqnarray}
\end{abstract}

\keywords{publications, bibliography --- astronomical databases:
  miscellaneous --- reference systems --- Galaxy: fundamental
  parameters --- Galaxy: kinematics and dynamics}

\section{Galactic rotation constants}

Calibration of the Galactic rotation curve is of crucial importance
for a range of open questions in Galactic astrophysics. For instance,
knowledge of the Galaxy's speed of rotation at the solar circle,
$\Theta_0$, allows derivation of the total mass of the Milky Way,
including its dark matter component, provided that we accurately know
the Sun's Galactocentric distance, $R_0$ as well as the {\it shape} of
the rotation curve. The latter would help us decompose the Milky Way's
mass contributions into disk, bulge, and dark or visible halo
components (e.g., Sofue et al. 2009; Xin \& Zheng 2013) and constrain
the local dark matter density (e.g., Salucci et al. 2010; Weber \& de
Boer 2010). In addition, access to an accurate Galactic rotation curve
is a fundamental stepping stone for Galactic distance determinations
based on tracer populations using radio observations (e.g., Reid et
al. 2014; Reid \& Dame 2016).

As a result, numerous studies have aimed at deriving the Galaxy's
rotation speed, either at the Sun's Galactocentric distance or as a
function of distance from the Galactic Center. Surprisingly, perhaps,
few comprehensive meta analyses have been undertaken to explore
intrinsic biases or systematic differences among $\Theta_0$ values
resulting from different tracer populations or from a range of
underlying assumptions. This is what we set out to do here.

In a series of recent papers (de Grijs et al. 2014; de Grijs \& Bono
2014, 2015, 2016; henceforth Papers I--IV), we explored whether a
number of key distances in the Local Group may have been subject to
publication bias, i.e., the so-called `bandwagon effect' where new
results are only published if they are at least somewhat consistent
with previously published values. In Paper IV, we added to our
previous analyses of the distance to the Magellanic Clouds (Papers I
and III) and the M31 group (Paper II) by carefully assessing the
distance to the Galactic Center. We concluded that the body of $R_0$
estimates available in the literature was not obviously skewed by
publication bias, and we derived a statistically well-justified value
of $R_0 = 8.3 \pm 0.2 \mbox{ (statistical)} \pm 0.4$ (systematic)
kpc. We will hence use $R_0 = 8.3$ kpc throughout the present paper as
our reference value.

In Paper IV, we also undertook a preliminary analysis of the behavior
of the $\Theta_0/R_0$ ratio as a function of publication date based on
a limited parameter set drawn from the literature we had perused to
compile our database of $R_0$ values. Given the prevailing
uncertainties affecting $R_0$ measurements, and in particular those
dominating the (more uncertain) $\Theta_0$ values, the $\Theta_0/R_0$
ratio is usually much better determined than the individual
measurements contributing to it. Our preliminary analysis of the
$\Theta_0/R_0$ ratio as a function of publication date led us to
conclude that between 1990 and 2007 the trend, if any, remained almost
flat, with $\langle \Theta_0/R_0 \rangle \approx 28$ km s$^{-1}$
kpc$^{-1}$, while it may since have increased, reaching $\langle
\Theta_0/R_0 \rangle \approx 32$ kms$^{-1}$ kpc$^{-1}$
(2013--2015). We consequently concluded that our recommended $R_0$
value thus implied that $\Theta_0$ should be substantially revised
upward.

In this paper, we go significantly beyond that initial exploration by
compiling the most comprehensive catalog of published $\Theta_0$
values available to date. We will clearly show that our conclusion
from Paper IV is only strictly valid for young tracer populations,
with a less obvious need for a significantly increased value for
$\Theta_0$ implied by the Galaxy's main mass components. In Section 2
we outline our approach to compiling our database of Galactic rotation
constants and discuss the catalog's overall properties. In Section 3
we consider subsamples selected on the basis of the tracers used to
derive $\Theta_0$ and explore their differences. Section 4 provides a
holistic overview of the statistical differences uncovered in the
paper, concluding with our main recommendation as regards the
statistically most appropriate values for the prevailing Galactic
rotation constants.

\section{Data mining the literature}

\begin{figure*}[ht!]
\epsscale{0.75}
\plotone{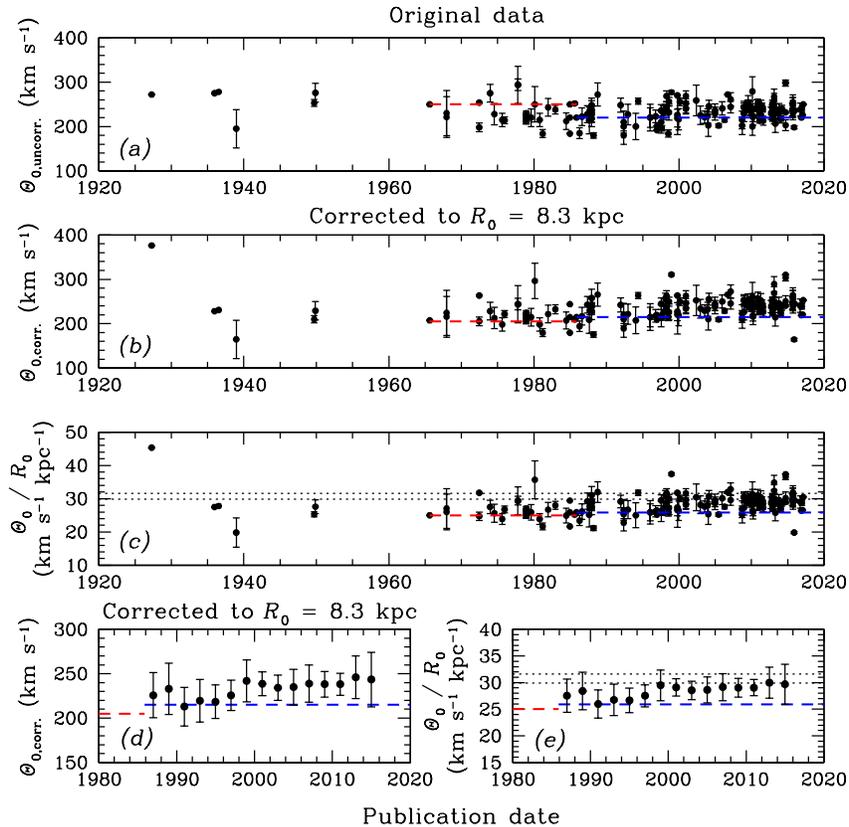}
\caption{Published Galactic rotation constants as a function of
  publication date. (a) Original $\Theta_0$ values, including their
  original 1$\sigma$ error bars. (b) $\Theta_0$ values corrected to a
  common solar Galactocentric distance of $R_0 = 8.3$ kpc (only for
  those measurements where this scaling is warranted), adopting the
  same error bars as in panel (a). (c) $\Theta_0/R_0$ ratio as a
  function of publication date. To determine the error bars, we simply
  scaled the original error bars for the $\Theta_0$ values, adopting
  the Galactocentric distances as fixed values. The horizontal dotted
  lines encompass the theoretically expected range derived by McMillan
  and Binney (2010). (d) and (e) Three-year averages based on
  independent, non-overlapping period ranges. The vertical `error
  bars' represent the distributions' Gaussian $\sigma$'s. The red and
  blue horizontal dashed lines (all panels) represent the IAU's
  recommendations from 1965 and 1985, respectively.}
\label{fig1}
\end{figure*}

Despite its importance as a means to determine the mass of the Milky
Way as a whole, few meta analyses of the Galactic rotation speed, at
the solar circle or otherwise---or of the better-determined ratio of
the rotation speed and the associated Galactocentric distance---have
been published. Perhaps the first comprehensive review of the
literature on this matter was provided by Kerr \& Lynden-Bell
(1986). This latter manuscript contains a summary of the deliberations
that led to the 1985 revision of the International Astronomical
Union's (IAU) recommended values for the Galactocentric distance, $R_0
= 8.5$ kpc, and the Galactic rotation speed at the solar circle,
$\Theta_0 = 220$ km s$^{-1}$, values that are still used for reference
today. The only more recent reviews of the Galactic rotation speed
were published by Sofue (2016) and Bland-Hawthorn \& Gerhard
(2016).\footnote{Clearly, most authors who derived new values of the
  relevant rotation constants compared their results with previously
  published determinations, but comprehensive reviews have been
  largely lacking since Kerr \& Lynden-Bell (1986); Bland-Hawthorn \&
  Gerhard (2016) do not consider any trends with publication date.}

We thus set out with a blank slate, and with the aim to uncover as
many determinations of $\Theta_0$, the solar circular velocity $v_{\rm
  c}$ (which includes a component $v_\odot$ in the direction of
Galactic rotation that must be corrected for to determine $\Theta_0$)
and the Oort constants $A$ (on its own) and/or $A$ and $B$. We
followed a similar approach as that used in Papers I--IV. First, we
searched the NASA/Astrophysics Data System (ADS) database for articles
referring to the Milky Way and containing one or more of the abstract
keywords `rotation curve,' any variety of the words `kinematic'
(kinematic, kinematics, kinematical) or `dynamics' (dynamics,
dynamical), or `Oort.' This resulted in an initial collection of 9,690
articles, spanning the period from the original dynamics papers by
Oort (1927a,b) until the end of 2017 June.

Next, we perused all articles in detail, looking for newly determined
or rederived Galactic rotation constants, while also following the
reference trail to previously published papers used as comparison
material for the newly (re-)derived rotation constants. Eventually,
this led to a compilation of 162 $\Theta_0$ values which were newly
obtained at the time of their publication. The NASA/ADS information
pages report that the database contains all articles published in the
main astrophysics journals since 1975, while the collection's
completeness of historical records is continuously increasing. Since
we will base our main conclusions in this paper on post-1985 data, the
most important factor affecting the completeness of our own database
is the question as to whether we have managed to track down all
relevant values in the literature. We are confident that we have found
the vast majority, but we call upon the community to submit additional
entries we may have missed for possible inclusion in our online
database.

As for our previous papers in this series, our database can be
accessed from http://astro-expat.info/Data/pubbias.html,\footnote{A
  permanent link to this page can be found at
  http://web.archive.org/web/20160610121625/http://astro-expat.info/Data/pubbias.html;
  members of the community are encouraged to send us updates or
  missing information, which will be included in the database if
  appropriate.} where we provide full bibliographic references and
direct links to the original articles. The uncertainties, where
available, are the statistical uncertainties only; no explicit
references to systematic uncertainties were found. However, one can of
course gain insight into the latter by examining the different values
provided by a number of authors for different assumptions made (e.g.,
Lynden-Bell \& Lin 1977; Einasto et al. 1979; Jackson 1985; Caldwell
\& Coulson 1988; Merrifield 1992; Miyamoto \& Zhu 1998; Feast et
al. 1998; Zhu 2000; Shen \& Zhu 2007; Xue et al. 2008; Yuan et
al. 2008; Reid et al. 2009a,b; Bovy et al. 2009; Bobylev 2013; Zhu \&
Shen 2013; Branham 2014; Bobylev \& Bajkova 2015).

Note that not every paper actually reports the relevant Galactic
rotation speed, or even $v_{\rm c}$. However, numerous articles report
the best-fitting Oort constants $A$, or both $A$ and $B$. The latter
are defined as
\begin{eqnarray}
A &=& \frac{1}{2} \left( \frac{\Theta_0}{R_0} - \frac{{\rm d}v}{{\rm
    d}r}|_{R_0} \right);\\ 
B &=& -\frac{1}{2} \left( \frac{\Theta_0}{R_0} + \frac{{\rm d}v}{{\rm
    d}r}|_{R_0} \right),
\end{eqnarray}
which allow for the presence of a velocity gradient, ${\rm d}v/{\rm
  d}r$, at the solar circle. For a flat rotation curve at $R_0$, this
simplifies to
\begin{eqnarray}
A &=& \frac{1}{2} \frac{\Theta_0}{R_0};\\ 
B &=& -\frac{1}{2} \frac{\Theta_0}{R_0},
\end{eqnarray}
so that 
\begin{equation}
\Theta_0 = 2AR_0 = (A-B)R_0.
\label{oort.eq}
\end{equation} 
Therefore, if $\Theta_0$ had not been determined directly but $A$ or
$A$ and $B$ was (were) available in a given paper, we used
Eq. (\ref{oort.eq}) to obtain an estimate of $\Theta_0$ under the
assumption of a flat rotation curve. Where we were required to proceed
in this manner, this has been indicated in the notes associated with
our final database.

We also included the Galactocentric distances $R_0$ adopted or derived
by their respective authors, given that the majority of the Galactic
rotation speeds included in our compilation are degenerate with
respect to the former. In the remainder of this paper, we will treat
these $R_0$ values as fixed, i.e., without considering any
uncertainties associated with their use (see Paper IV for a proper
treatment of these uncertainties). Only four of the Galactic rotation
speeds contained in our database are not directly scalable with $R_0$
(Lynden-Bell \& Frenk 1981; McCutcheon et al. 1983; Alvarez et
al. 1990; Frinchaboy \& Majewski 2005); these values have been omitted
in our subsequent analysis, although we note that most of their values
(with the exception of the low $\Theta_0$ reported by McCutcheon et
al. 1983) are largely consistent, at least within the uncertainties,
with the bulk of the determinations in our database, as well as with
our final recommendation.

Figure \ref{fig1}a shows our full data set of Galactic rotation speeds
as published in their original papers. In Fig. \ref{fig1}b, we have
homogenized the $\Theta_0$ values by scaling to a common
Galactocentric distance of $R_0 = 8.3$ kpc, our recommendation
resulting from the statistical analysis in Paper IV. Next, in
Fig. \ref{fig1}c, we show the $\Theta_0 / R_0$ ratios (for which the
actual value of $R_0$ adopted is unimportant given the degeneracy
between $\Theta_0$ and $R_0$). All panels also include indications of
the prevailing IAU recommendations from 1965 ($\Theta_0 = 250$ km
s$^{-1}$; $R_0 = 10$ kpc) and 1985 (red and blue horizontal dashed
lines, respectively). Finally, Figs \ref{fig1}d and e show, for the
post-1985 era, the successive three-year averages for $\Theta_0$
(rescaled to $R_0 = 8.3$ kpc) and $\Theta_0 / R_0$, respectively.
Panels (c) and (e) also include the theoretically predicted range for
$\Theta_0 / R_0$ from McMillan \& Binney (2010), which seems somewhat
overestimated, even in view of the multi-year averages shown in
Fig. \ref{fig1}e. In this context, one should keep in mind that the
vertical error bars shown for the multi-year averages are the Gaussian
widths, $\sigma$, of the distribution in a given period range; the
uncertainties on the mean values would be smaller by a factor of
$\sqrt{N}$ (where $N$ corresponds to the number of values contributing
to the mean) if our assumption of Gaussian distributions is
reasonable. Figures \ref{fig1}d and e suggest that any trend would
have flattened out around the year 2000 already. The Gaussian
$\sigma$'s do not seem to be correlated over this period, however, so
that we cannot claim to have uncovered publication bias or a
`bandwagon effect' since the turn of the last Century.

\section{Differences among tracer populations}

\begin{figure*}[ht!]
\epsscale{0.77}
\plotone{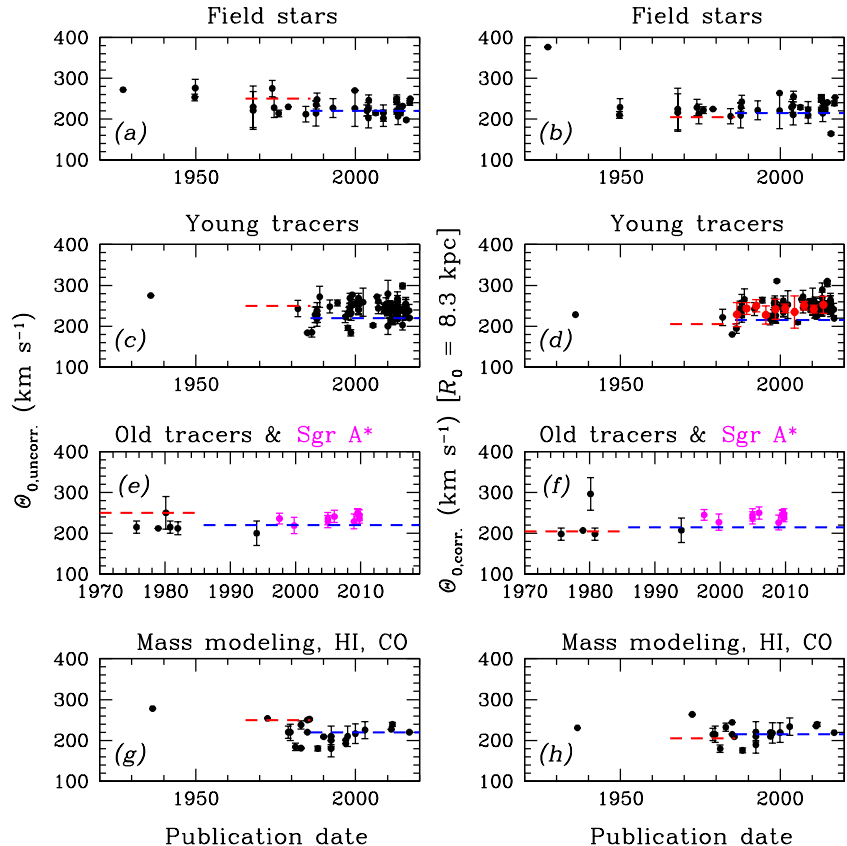}
\caption{As Fig. \ref{fig1}, but for different tracers separately. The
  left- and right-hand panels display the originally published
  measurements and their counterparts corrected to a common
  Galactocentric distance of $R_0 = 8.3$ kpc, respectively. The red
  data points in panel (d) represent the young tracers' four-year
  running average $\Theta_0$ value, again using independent successive
  period ranges.}
\label{fig2}
\end{figure*}

Figure \ref{fig2} shows (left) the originally published and (right)
the corrected Galactic rotation speeds as a function of publication
date and separated by stellar population tracer. We grouped the
entries in our database into `field stars,' young tracer populations,
rotation speeds based on the kinematics (proper motions) of Sgr A*
near the Galactic Center, and rotation speeds based on Galactic mass
modeling, H{\sc i} (neutral hydrogen), and carbon monoxide (CO) radio
observations. Note that, strictly speaking, the Sgr A* measurements as
well as many of the `young tracers' values are based on maser
astrometry, so that we could, in principle, have taken these
measurements together. As we will see below, the results for both
subgroups of measurements are fully consistent with each other.

We have collected a sufficient number of data points for the young
tracers to attempt the construction of a multi-year running
average. The resulting trend is overplotted in Fig. \ref{fig2}d, where
the red data points show the mean values and standard deviations of
the distribution, in successive bins of four years each. The four-year
averages from 1985 onward do not suggest any obvious trend within the
random fluctuations, which are mostly owing to small-number statistics
rather than to systematic uncertainties.

Among the field stars, we have included disk, bulge, bar, and halo
stars, as well as large samples of giant, main-sequence, and dwarf
stars. As such, the `field' population appears truly representative of
the Galaxy's stellar populations as a whole. Similarly, the $\Theta_0$
results based on Galactic mass modeling and those relying on H{\sc i}
and CO data trace the Galaxy's overall mass distribution. A cursory
examination of the right-hand panels of Fig. \ref{fig2} already tells
us that while the field stars, and the Galactic mass models, H{\sc i},
and CO observations are largely consistent with the prevailing IAU
recommendation (see the blue dashed lines), the young tracers
(composed of OB and supergiant stars, masers, Cepheid variables, H{\sc
  ii} regions, and young open clusters,\footnote{Although open
  clusters in the Galactic disk cover a range in ages, the $\Theta_0$
  measurements included in our database are based on samples of young
  clusters, with ages of up to 50 Myr (Shen \& Zhu 2007; Zhu 2009; Zhu
  \& Shen 2013).} as well as those values resulting from Sgr A*
kinematic measurements) are characterized by higher rotation
speeds. We will return to these differences in the next
section. (Since the few $\Theta_0$ measurements we have based on old
tracers were predominantly published before 1985, with a singke
exception, we will not include these values in our analysis.)

\begin{figure}[ht!]
\epsscale{1.2}
\plotone{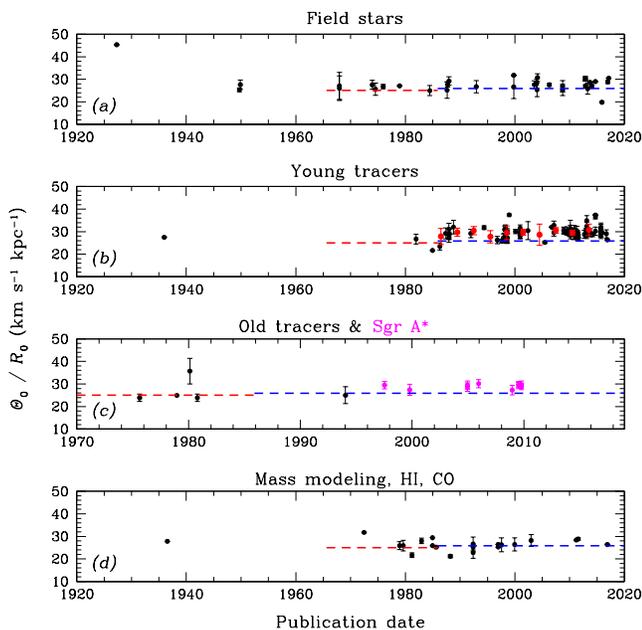}
\caption{As Fig. \ref{fig2}, but for the $\Theta_0 / R_0$ ratios. The
  red data points in panel (b) represent the young tracers' four-year
  running average.}
\label{fig3}
\end{figure}

A similar scenario is sketched by Fig. \ref{fig3}, which displays the
$\Theta_0 / R_0$ ratio as a function of publication date for the same
tracer populations. Again, the Galactic mass models and the results
from the H{\sc i} and CO observations and the field stars are largely
consistent with the prevailing IAU recommendation. The young
populations, as well as the Sgr A* kinematics, imply a significantly
larger ratio. The mean levels for each subpopulation, the
uncertainties on the means---formally defined by the Gaussian width of
the tracer distribution divided by the square root of the number of
measurements, $\sigma/\sqrt{N}$---and the overall $\sigma$'s of the
tracer populations are included in Table \ref{table1} for easy
quantitative comparison. We will return to a statistical discussion of
these values in the next section.

\begin{table*}
\caption{Quantitative comparison of the $\Theta_0$ and $\Theta_0/R_0$
  values for different tracers and different periods. $N$: Number of
  measurements.}
\begin{center}
\label{table1}
\begin{tabular}{@{}llcrcccr@{}}
\hline \hline
\multicolumn{1}{c}{Tracer} & \multicolumn{1}{c}{Period} & \multicolumn{2}{c}{$\Theta_0$ 
(km s$^{-1}$)} && \multicolumn{2}{c}{$\Theta_0/R_0$ (km s$^{-1}$ kpc$^{-1}$)} & \multicolumn{1}{c}{$N$} \\
\cline{3-4}\cline{6-7}
 & & \multicolumn{1}{c}{Mean} & \multicolumn{1}{c}{$\sigma$} && \multicolumn{1}{c}{Mean} & \multicolumn{1}{c}{$\sigma$} \\
\hline
Field stars   & 1965--1985    & $219.2 \pm  2.9$ &  7.6 && $26.42 \pm 0.35$ & 0.92 &   7 \\
              & 1985--present & $229.9 \pm  4.0$ & 20.2 && $27.70 \pm 0.49$ & 2.46 &  25 \\
Young tracers & 1985--present & $248.5 \pm  2.6$ & 19.8 && $29.94 \pm 0.32$ & 2.42 &  59 \\
Old tracers   & All           & $212.1 \pm 16.5$ & 43.7 && $25.55 \pm 2.01$ & 5.33 &   7 \\
Sgr A*        & All           & $240.8 \pm  2.6$ &  8.2 && $29.01 \pm 0.32$ & 1.00 &  10 \\
Mass models   & 1985--present & $219.4 \pm  3.5$ & 13.8 && $26.44 \pm 0.42$ & 1.69 &  16 \\
All           & 1965--1985    & $219.4 \pm  4.0$ & 22.1 && $23.44 \pm 0.48$ & 2.70 &  31 \\
              & 1985--present & $238.6 \pm  2.0$ & 22.0 && $28.74 \pm 0.24$ & 2.68 & 126 \\
\hline \hline
\end{tabular}
\end{center}
\end{table*}

\section{Is there a need for reassessment of the IAU recommendation?}

\begin{figure*}[ht!]
\plottwo{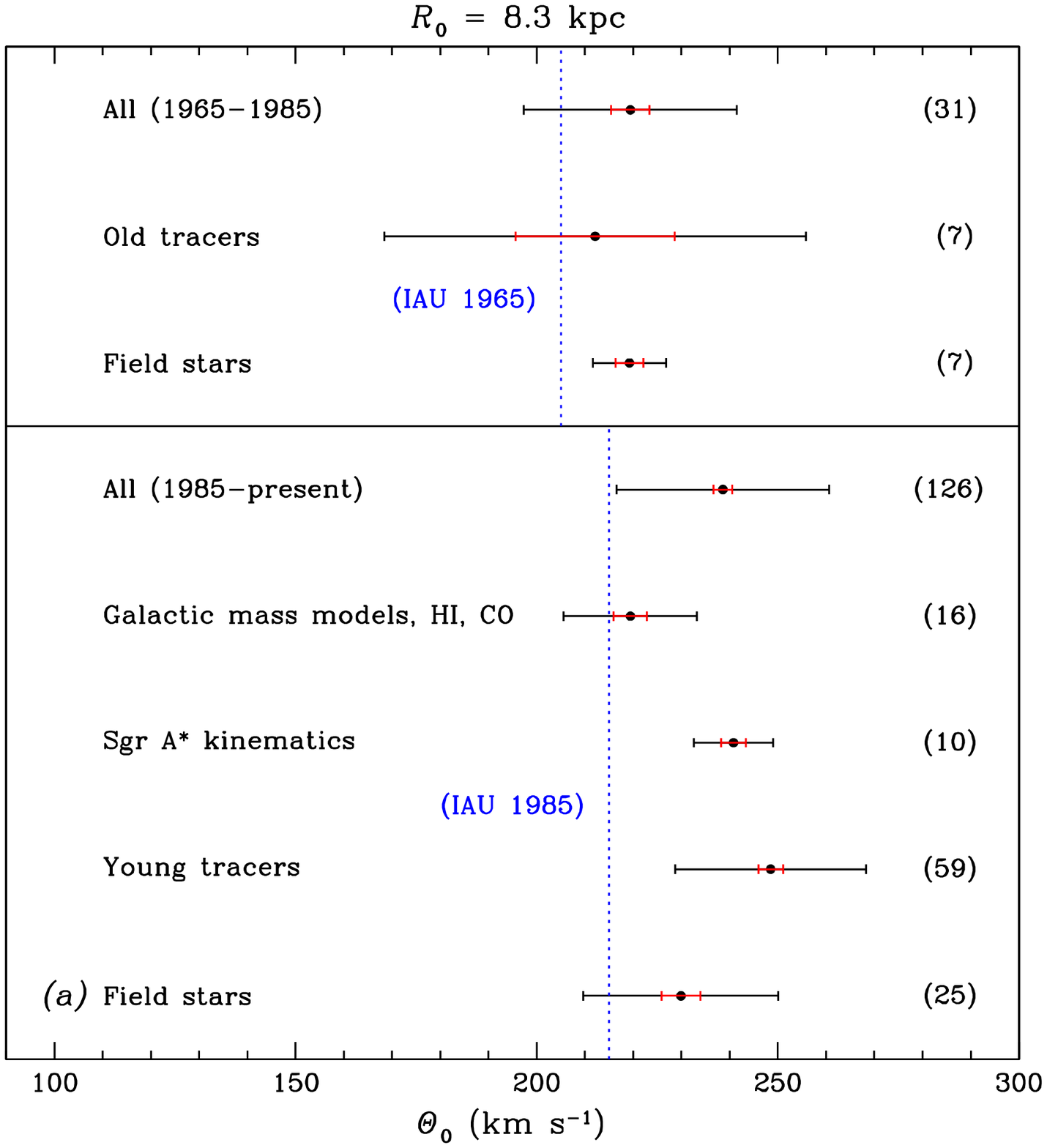} {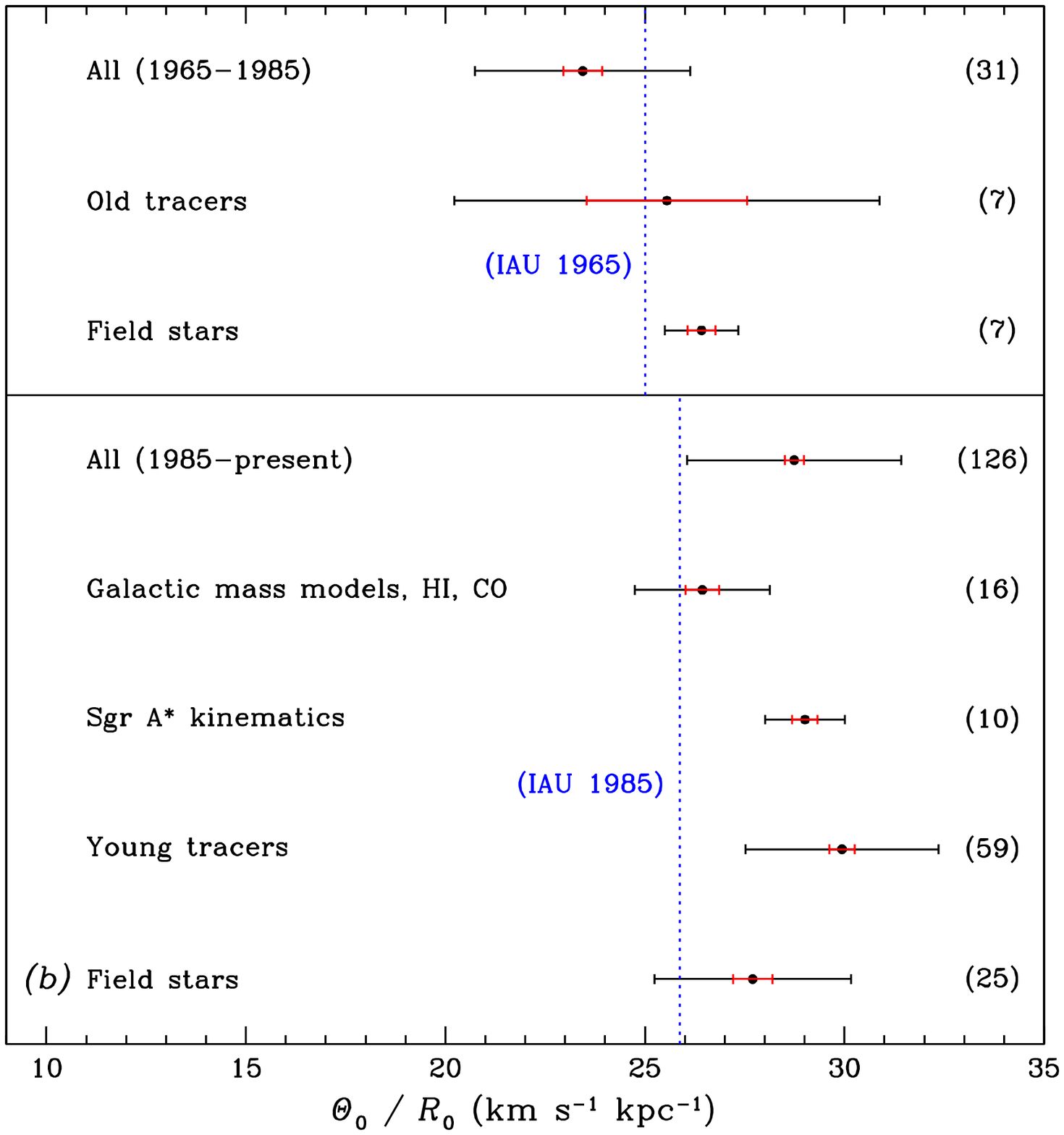}
\caption{Comparisons of (a) the $\Theta_0$ values (for $R_0 = 8.3$
  kpc) and (b) the $\Theta_0 / R_0$ ratios among the different tracers
  used. The top (bottom) parts of both panels largely refer to
  measurements from the 1965--1985 (1985--present) period (see Table
  \ref{table1} for details). The prevailing IAU recommendations are
  shown as vertical blue dotted lines. The inner, red error bars
  represent the uncertainties in the tracers' mean values, defined as
  $\sigma/\sqrt{N}$ for $N$ measurements (shown in brackets), while
  the outer, black error bars encompass the Gaussian $\sigma$'s of
  their distributions.}
\label{fig4}
\end{figure*}

Figure \ref{fig4} is a graphical representation of the Galactic
rotation speed at the solar circle and the $\Theta_0/R_0$ ratio
implied by the different tracer populations. For ease of comparison,
we have also included the IAU-recommended values using vertical blue
dotted lines. When considering the post-1985 values, the impression we
gained in the previous section is solidified: the young tracers as
well as the Sgr A* kinematics imply a much higher rotation speed and a
larger $\Theta_0/R_0$ ratio than the old tracers (although note that
the latter measurements are few in number and mostly published before
1985), the field stellar population, radio observations of H{\sc i}
and CO gas kinematics, and Galactic mass modeling.

In order to provide further support for the reality of statistical
differences among the tracer populations, we performed a statistical
Kruskal--Wallis $H$ test. This is a non-parametric test which is used
to determine whether three or more groups are statistically the same
in terms of their mean {\it ranks}. The test does not assume that the
data points are distributed according to a normal distribution, so it
is more robust than the one-way ANOVA test. The underlying null
hypothesis is that all groups have similar mean ranks.

We applied the Kruskal--Wallis test to the 25 post-1985 $\Theta_0$
values for the field stars, the 59 post-1985 values for the young
tracers, all 10 $\Theta_0$ values based on the proper motion of Sgr
A*, and the 16 post-1985 data points resulting from mass modeling. We
first sorted all values, irrespective of their provenance, in
ascending order. Next, we assigned ranks to the sorted values,
assigning the average rank to any tied values. Specifically, the mean
ranks for the post-1985 data composed of field stars, young tracers,
Sgr A*-based values, and those based on mass modeling are,
respectively, 51.1, 70.1, 66.0, and 33.7. We then added up the
different ranks for each group and calculated the $H$ statistic:

\begin{equation}
H = \Bigl( \frac{12}{N(N-1)} \sum_{j=1}^c \frac{T_j^2}{N_j} \Bigr) - 3(N+1),
\end{equation}
where $N = 110$ is the total number of data points, $c$ is the number
of data points in a group, $T_j$ is the sum of the ranks in the
$j^{\rm th}$ group, and $n_j$ is the size of the $j^{\rm th}$
group. For our specific data set, $H = 16.58$.

This $H$ value must be compared with the critical $\chi^2$ value for
$c-1$ degrees of freedom and a nominal $p$ value of 0.05 (which is
$\chi^2_{\rm crit} = 7.82$). If the critical $\chi^2$ value is less
than the $H$ statistic, we can reject the null hypothesis that the
medians of the different groups are equal. This is indeed the case for
the four groups considered here. The corresponding $p$ value is
0.00086.

Curiously, few authors have commented on the significantly larger
values they obtained for either $\Theta_0$ or $\Theta_0/R_0$ based on
young tracers compared with field stars or radio data. In fact, most
authors reporting higher-than-average values selectively compared
their results with previously published values for similar populations
(e.g., Elias et al. 2006). This could be considered selective
publication bias.

Let us, instead, take a holistic view and consider the underlying
causes of these higher values resulting from young tracers. While the
field stars and the H{\sc i}/CO distribution represent a smooth
three-dimensional mass distribution, the young tracers are
predominantly found in the spiral arms, mostly near the solar circle,
where non-circular motions---including vertical oscillations (e.g.,
Kerr \& Lynden-Bell 1986; Bobylev \& Bajkova 2015) or radial motions
toward the Galactic Center (but see Reid et al. 2009b for an opposing
view)---may be significant (e.g., Foster \& Cooper 2010). In addition,
these young tracer populations tend to be located in a much thinner
disk-like structure than, e.g., the K and M giants (Zhu 2000; Huang et
al. 2016) or the Southern Proper Motion Program (M\'endez et
al. 1999), the Tycho-2 (Olling \& Dehnen 2003), SDSS (Sirko et
al. 2004; Xue et al. 2008; Sch\"onrich 2012), APOGEE (Bovy et
al. 2012; Bovy 2014), RAVE (Sharma et al. 2014), or LAMOST stars
(Huang et al. 2016) making up the diverse field stellar population in
the Galaxy's disk and halo.

It has also been suggested that the different Galactic rotation speeds
determined on the basis of local ($< 1$ kpc) with respect to more
distant Cepheid variables could be caused by an intrinsic kinematic
bias affecting the local sample, which may be characterized by its own
local kinematics instead of the overall Galactic rotation properties
(Glushkova et al. 1999), possibly explaining a local dip in the
Galactic rotation curve. However, independent estimates based on both
{\sl Hipparcos} and Southern Proper Motion Catalog (Platais et
al. 1998) stars yielding mutually consistent rotation speeds seem to
have al but invalidated those suggestions; comprehensive and
up-to-date Galactic rotation curves no longer show any dip near the
Sun's Galactocentric distance, at least not those based on the main
mass tracers (see, e.g., Bland-Hawthorn \& Gerhard 2016). It is still
possible that the appearance of such a dip could be limited to the
young(est) populations located in the Perseus spiral arm (e.g., Sofue
2013; Genovali et al. 2014). Genovali et al. (2014) also suggested
that the clumpy distribution of classical Cepheids across the Galactic
thin disk and their possible association with giant molecular clouds
may hamper their use for Galactic rotation measurements.

Note that the H{\sc i}, CO, and field population tracers also imply
larger values for both $\Theta_0$ (at $R_0 = 8.3$ kpc) and the
$\Theta_0/R_0$ ratio than the canonical IAU recommendation, although
the widths of their full distributions encompass the IAU values within
their 1$\sigma$ ranges. In view of these distributions, and
considering that both the field populations and the H{\sc i} or CO gas
kinematics reflect the underlying Galactic mass distribution most
closely, we contend that the post-1985 combination of these
measurements may indeed yield the most appropriate value for the
Galactic rotation speed for $R_0 = 8.3$ kpc.

The 40 post-1985 measurements of $\Theta_0|_{R_0 = 8.3\; {\rm kpc}}$
based on field star samples, H{\sc i} or CO observations, and Galactic
mass modeling yield $\Theta_0 = 225.1 \pm 3.2 \mbox{ km s}^{-1}
(\sigma = 20.3 \mbox{ km s}^{-1})$. Note that, while dominated by a
statistical component, the uncertainty does include a contribution
from a systematic component, given that the contributing $\Theta_0$
values include such uncertainties. Similarly, the resulting
$\Theta_0/R_0$ ratio based on these values is $\Theta_0 / R_0 = 27.12
\pm 0.39 \mbox{ km s}^{-1} \mbox{ kpc}^{-1} (\sigma = 2.45 \mbox{ km
  s}^{-1} \mbox{ kpc}^{-1})$.

At this point, it would be wise to perform a final check of the
Galactic rotation speed at the solar circle by considering the
Galaxy's rotation curve with respect to the external reference frame
provided by the Milky Way's satellite galaxies and other Local Group
members, as well as the Milky Way's GD-1 giant stream. Fortunately, a
number of studies have attempted to do just this. Lynden-Bell \& Lin's
(1977) and Einasto et al.'s (1979) original attempts yielded rotation
speeds that bracketed the expected range. Lynden-Bell \& Lin (1977)
determined $\Theta_0 = 244 \pm 42$ km s$^{-1}$ for $R_0 = 8.3$ kpc
(with a reduced uncertainty of 13 km s$^{-1}$ resulting from the
application of more a priori constraints), while Einasto et al. (1979)
found $\Theta_0 = 215 \pm 7$ km s$^{-1}$, again for $R_0 = 8.3$
kpc. More recent derivations of $\Theta_0$ in external reference
frames, all recalibrated for $R_0 = 8.3$ kpc, have resulted in
$\Theta_0 = 215 \pm 22$ (30) km s$^{-1}$ (Kochanek et al. 1996) and
$\Theta_0 = 219 \pm 13$ ($216 \pm 18$) km s$^{-1}$ (Koposov et
al. 2010).

This thus implies that there is no compelling evidence suggesting that
the IAU-recommended $\Theta_0 = 220$ km s$^{-1}$ is no longer valid,
although it would be more appropriate to increase the recommendation
to $\Theta_0 = 225$ km s$^{-1}$, {\it provided} that the recommended
Galactocentric distance is reduced from $R_0 = 8.5$ kpc to $R_0 = 8.3$
kpc (Paper IV). A comparison of the Galactic rotation speeds implied
by the Galactic mass tracers with those resulting from considering an
external reference frame implies a systematic uncertainty of order
$\pm 10$ km s$^{-1}$. Combined with the results from our statistical
analysis in Paper IV, we thus conclude that the most appropriate set
of Galactic rotation constants is
\begin{eqnarray}
\Theta_0 &=& 225 \pm 3 \mbox{ (statistical)} \pm 10 \mbox{ (systematic)
  km s}^{-1} \nonumber \\
~\\
R_0 &=& 8.3 \pm 0.2 \mbox{ (statistical)} \pm 0.4 \mbox{ (systematic)
  kpc},\nonumber
\end{eqnarray}
so that
\begin{eqnarray}
\Theta_0 / R_0 = 27.12 &\pm& 0.39 \mbox{ (statistical)} \\
&\pm& 1.78 \mbox{ (systematic) km s}^{-1} \mbox{ kpc}^{-1}.\nonumber
\end{eqnarray}

\section*{Acknowledgements}

R. d. G. is grateful for research support from the National Natural
Science Foundation of China through grants 11373010, 11633005, and
U1631102. This work was also partially supported by PRIN-MIUR
(2010LY5N2T), `Chemical and dynamical evolution of the Milky Way and
Local Group galaxies' (PI F. Matteucci). This research has made
extensive use of NASA's Astrophysics Data System Abstract Service.

\end{document}